\documentclass[final,3p,times,twocolumn]{elsarticle}

\usepackage{amssymb}
\usepackage{amsmath,amssymb,exscale}
\usepackage{graphicx}
\usepackage{epsfig}
\usepackage{multicol}
\usepackage{color}
\usepackage{xcolor}
\usepackage{hyperref}
\usepackage{mathrsfs}
\usepackage{hyperref}
\hypersetup{colorlinks,bookmarksopen,bookmarksnumbered,citecolor=rossos,
linkcolor=black,pdfstartview=FitH,urlcolor=blus}
\usepackage{amsfonts}
\usepackage{slashed}
\usepackage{blindtext}
 \usepackage{fancyhdr}
\usepackage{hyperref}
\usepackage{mathtools}
\biboptions{numbers,sort&compress}

\allowdisplaybreaks

\definecolor{blus}{cmyk}{1,1,0,0.}
\definecolor{verdes}{cmyk}{0.99,0,0.99,0.02}
\definecolor{rossos}{cmyk}{0,1,1,0.55}
\definecolor{greeny}{cmyk}{0.99,0,0.59,0.98}
\definecolor{redy}{cmyk}{0,1,1,0.40}

\newcommand{\bp}{M_P}
\def\Lag{\mathscr{L}}

\def\be{\begin{equation}}
\def\ee{\end{equation}}
\def\bea{\begin{eqnarray}}
\def\eea{\end{eqnarray}}

\def\ba{\begin{array} }
\def\ea{\end{array}}
\def\bac{\begin{array} {c}}
\def\bacc{\begin{array} {cc}}
\def\baccc{\begin{array} {ccc}}

\def\psl{\hbox{\hbox{${p}$}}\kern-1.9mm{\hbox{${/}$}}}
\def\dsl{\hbox{\hbox{${\partial}$}}\kern-1.7mm{\hbox{${/}$}}}
\def\Dsl{\hbox{\hbox{${D}$}}\kern-2.1mm{\hbox{${/}$}}}

\definecolor{red}{rgb}{1,0,0}

\def\hhref#1{\href{http://arxiv.org/abs/#1}{arXiv:#1}}  
\hypersetup{colorlinks=true, linkcolor=red}
 
\journal{the arXiv}

\begin{document}

\begin{frontmatter}

\title{\vspace{-2cm}\huge {\color{redy} Pulsar Timing Arrays and  \\  Primordial Black Holes from a Supercooled Phase Transition
}}

\author{\vspace{1cm}{\large {\bf Alberto Salvio}}}

\address{\normalsize \vspace{0.2cm}Physics Department, University of Rome and INFN Tor Vergata, Italy \\

\vspace{0.3cm} 
 \vspace{-1cm}
 }

\begin{abstract}
\noindent An explicit realistic model featuring a supercooled phase transition,  which allows us to explain the background of gravitational waves recently detected by pulsar timing arrays, is constructed. In this model the phase transition corresponds to radiative symmetry breaking (and mass generation) in a dark sector featuring  a dark photon associated with the broken symmetry. The completion of the transition is ensured by a non-minimal coupling between gravity and the order parameter and fast reheating occurs thanks to a preheating phase.  Finally, it is also shown that the model leads to primordial black hole production.

\end{abstract}


\end{frontmatter}
 
 \begingroup
\hypersetup{linkcolor=blus}
\tableofcontents
\endgroup


\section{Introduction}\label{introduction}
  
  \vspace{-0.3cm}

 Gravitational wave (GW) astronomy has become an extremely active and exciting field of physics after the discovery of GWs from  binary black-hole and neutron-star mergers~\cite{Abbott:2016blz,TheLIGOScientific:2016wyq1,LIGOScientific:2017ync}. More recently, the interest in this field has been further increased by the detection of a background of GWs by pulsar timing arrays (PTAs), including the North American Nanohertz Observatory for Gravitational Waves (NANOGrav), the Chinese Pulsar Timing Array (CPTA), the European Pulsar Timing Array (EPTA) and the Parkes Pulsar Timing Array (PPTA)~\cite{NANOGrav:2023gor,Antoniadis:2023ott,Reardon:2023gzh,Xu:2023wog}. 

 Among several possibilities to interpret the PTA background, an interesting option is  first-order phase transitions (PTs). Indeed, these phenomena are not present in the Standard Model (SM) and, therefore, the detection of a GW background generated in this way (see Ref.~\cite{Maggiore:2018sht} for a textbook introduction) would be clear evidence for new physics.
A PT interpretation of the signals detected by PTAs performed by the NANOgrav collaboration and relevant for the present study was provided in Ref.~\cite{NANOGrav:2023hvm} and other independent discussions of PT interpretations were given e.g.~in Refs.~\cite{Ashoorioon:2022raz,Antoniadis:2023zhi,Bringmann:2023opz,Madge:2023cak,Zu:2023olm,Han:2023olf,Fujikura:2023lkn,Kitajima:2023cek,Bai:2023cqj,Addazi:2023jvg,Athron:2023mer,Lu:2023mcz,Xiao:2023dbb,Li:2023bxy,Ghosh:2023aum,Figueroa:2023zhu,Wu:2023hsa,DiBari:2023upq,Gouttenoire:2023bqy,Salvio:2023ynn,He:2023ado,Ellis:2023oxs,Chen:2023bms}.

 First-order PTs generally occur when the symmetries are predominantly broken (and masses are then generated) through radiative corrections~\cite{Coleman:1973jx,Gildener:1976ih,Witten:1980ez,Salvio:2023qgb} (see Ref.~\cite{Salvio:2020axm} for a review). This radiative symmetry breaking (RSB) is  interesting also because it has the unique property of allowing a perturbative (and thus calculable) description of PTs~\cite{Salvio:2023qgb}. Another distinctive property of such PTs is the presence of a period of supercooling when the temperature dropped much below the critical temperature~\cite{Witten:1980ez,Salvio:2023qgb}. 

  The  GW background detected by PTAs could be due to a supercooled PT associated with an RSB at a scale around the GeV~\cite{Gouttenoire:2023bqy,Salvio:2023ynn}. However, so far a concrete model with these characteristics has not been found and, therefore, its existence has not been guaranteed. Indeed, as pointed out in~\cite{Athron:2023mer}, identifying such a model is very challenging.

 {\it As a proof of concept, the purpose of this paper is to solve this issue by constructing an explicit realistic model that can explain the PTA background through a supercooled PT associated with RSB.}

 Besides generating a GW background, supercooled PT associated with an RSB also naturally lead to primordial black hole (PBH) production through the ``late-blooming mechanism" (see e.g.~Refs.~\cite{Kodama:1982sf,Liu:2021svg,Hashino:2021qoq,Kawana:2022olo,Gouttenoire:2023naa} for a quantitative introduction):  since the false vacuum decay of PTs is statistical for quantum and thermal reasons, in distinct patches the transition can occur at different times. Patches where the transition occurs the latest undergo the longest vacuum-dominated stage and, therefore, develop large over-densities,  which collapse into PBHs. These PBHs can then account for a fraction $f_{\rm {PBH}}$ of the dark matter density. 

 {\it Another purpose of the present work is to identify regions of the parameter space of the above-mentioned model where the PBH production is significant and the PTA detection is explained at the same time.  In the following sections we show that all this is possible and in the last section we highlight the main results achieved.}

  \vspace{-0.1cm}
 
\section{A minimal model} \label{A minimal model}

  \vspace{-0.2cm}
  
Let us start by considering a classically scale-invariant dark sector with RSB at the GeV scale. The simplest option is the Coleman-Weinberg construction~\cite{Coleman:1973jx}, which features an Abelian gauge symmetry, $U(1)_D$, and a charged scalar field $\phi$. The corresponding no-scale Lagrangian is given by
\be \Lag^{\rm ns} = - \frac14 \mathcal{A}_{\mu\nu}\mathcal{A}^{\mu\nu} +(D_\mu \phi)^\dagger \, D^\mu \phi  
- \frac1{4}\lambda_\chi\chi^4,\ee 
where $\mathcal{A}_{\mu\nu}\equiv \partial_\mu \mathcal{A}_\nu-\partial_\nu\mathcal{A}_\mu$ (the vector $\mathcal{A}_\mu$ is the $U(1)_D$ gauge boson),
 $\chi\equiv\sqrt2|\phi|$ and $D_\mu\phi=(\partial_\mu+ie_d\mathcal{A}_\mu)\phi$ (the parameter $e_d$ is the $U(1)_D$ gauge coupling). The symmetry $U(1)_D$ is the one that undergoes RSB. This happens through a radiatively induced VEV of $\chi$ at the required energy scale, $\chi_0\sim~$GeV. As well known, RSB takes place because quantum corrections produce the following one-loop effective potential for $\chi$
\be V_q(\chi) = \frac{\bar \beta}4\left(\log\frac{\chi}{\chi_0}-\frac14\right)\chi^4,\label{CWpot}\ee
where
\be \bar\beta \equiv \left[\mu\frac{d\lambda_{\chi}}{d\mu} \right]_{\mu=\tilde\mu} >0 \ee
and $\tilde\mu$ is a value of the renormalization group scale $\mu$ at which $\lambda_\chi = 0$. In words, $\bar\beta$ is the $\beta$ function of the quartic coupling at the energy scale  where such coupling vanishes.

However, the dark sector here interacts with the SM. Besides gravity, which is very weak, there are two possible operators that can mediate such interactions. One is the Higgs portal   
 \be  \Lag_{\chi h} = \frac12 \lambda_{\chi h} \chi^2 |{\mathcal H}|^2,\label{portal} \ee
where ${\mathcal H}$ is the SM Higgs doublet and $\lambda_{\chi h}$ is the corresponding quartic coupling. The other one is a kinetic mixing between the SM photon, $\mathcal{F}_\mu$, and $\mathcal{A}_\mu$:
\be\Lag_{\rm mix} = -\frac{\eta}{2}\mathcal{A}_{\mu\nu}\mathcal{F}^{\mu\nu}, \label{mixL}\ee
where $\mathcal{F}_{\mu\nu}\equiv \partial_\mu \mathcal{F}_\nu-\partial_\nu\mathcal{F}_\mu$ 
and $\eta$ is the mixing parameter.  

In general we can also introduce a non-minimal coupling $\xi_\chi$ appearing in the Lagrangian density through the term 
\be -\frac12\xi_\chi\chi^2R, \label{non-min}\ee
where $R$ is the Ricci scalar. The pure-gravity Lagrangian is assumed to be the standard Einstein-Hilbert one  at the energies that are
relevant for this work. Note that the term in~(\ref{non-min}) does not lead to an appreciable modification of gravity in our case because $\chi_0\sim~$GeV$~\ll \bp$, unless one takes $\xi_\chi$ extremely large, which we do not. However, as will be shown in Sec.~\ref{The (improved) supercool expansion}, that term  can ensure that the PT completes.

The mixing in~(\ref{mixL}) can be removed by performing the field redefinition $\{\mathcal{A}_\mu,\mathcal{F}_\mu\}\to\{A'_\mu,A_\mu\}$ given by
\be \left(
\begin{array}{c}
 \mathcal{A}_\mu  \\
\mathcal{F}_\mu  \\
\end{array}
\right)= \left(
\begin{array}{cc}
 \frac{1}{\sqrt{1-\eta ^2}} & 0 \\
 -\frac{\eta }{\sqrt{1-\eta ^2}} & 1 \\
\end{array}
\right)
\left(
\begin{array}{c}
 A'_\mu  \\
A_\mu  \\
\end{array}
\right).\ee
However, after this field redefinition, $\eta$ still appears in the Lagrangian because 
\bea
(D_\mu \phi)^\dagger \, D^\mu \phi 
=
(\partial_\mu \phi)^\dagger \, \partial^\mu \phi+\frac{e_d^2 \chi^2  A'_\mu A'^\mu}{2(1-\eta^2)} +\frac{e_d J'_\mu A'^\mu}{\sqrt{1-\eta^2}}, 
\label{Kinphi} \eea
with $J'_\mu\equiv i\left[\left(\partial_\mu\phi\right)^\dagger\phi-\phi^\dagger \partial_\mu \phi\right]$,
and because of the interaction term $e J_\mu\mathcal{F}^\mu$ between $\mathcal{F}_\mu$ and the SM electromagnetic current $J_\mu$,
 i.e. 
\be e J_\mu\mathcal{F}^\mu = e J_\mu A^\mu -\frac{e\eta J_\mu A'^\mu}{\sqrt{1-\eta^2}}. \label{EMcurrentInt} \ee

The second term in the right-hand side of~(\ref{Kinphi}) leads to a background-dependent squared  mass of the ``dark photon" $A'_\mu$, namely $m_d^2(\chi) = e_d^2 \chi^2/(1-\eta^2).$
After RSB $\chi$ acquires a VEV $\chi_0$ and this leads to the dark photon mass $m_d(\chi_0)$. Since in our current setup $\chi_0$ is much smaller than the EW scale and we demand the validity of perturbation theory ($e_d$ must not be large), $m_d(\chi_0)$ should be much smaller than the EW scale too.

The second term in the right-hand side of~(\ref{EMcurrentInt}) is instead an interaction  between $J_\mu$ and $A'_\mu$. Therefore, in order to satisfy the experimental limits $\eta$ should be small (see~\cite{Fabbrichesi:2020wbt,Graham:2021ggy} for reviews).  So 
\be \hspace{-0.14cm} m_d^2(\chi) \approx e_d^2 \chi^2, ~~ \frac{J'_\mu A'^\mu}{\sqrt{1-\eta^2}} \approx J'_\mu A'^\mu, ~~  \frac{\eta J_\mu A'^\mu}{\sqrt{1-\eta^2}}\approx \eta J_\mu A'^\mu  \label{mdIntApSM}\ee
and the dark photon is coupled to the electromagnetic current through a millicharge $e\eta$. The Higgs portal coupling $\lambda _{\chi h}$ should also be very small because the mass $m_\chi$ of $\delta\chi \equiv\chi-\chi_0$ is also much smaller than the EW scale: $m_\chi = \sqrt{\beta_\chi}\chi_0\ll \chi_0\sim~$GeV. 
Note that the smallness of $\lambda _{\chi h}$ and $\eta$ (and of gravity) ensures that, barring quantum corrections, the scale invariance of the dark sector is only mildly broken by its interactions.

  \section{The (improved) supercool expansion}\label{The (improved) supercool expansion}
  
  As recalled in the introduction, in a PT associated with RSB supercooling always occur. The field $\chi$ is trapped in the false minimum, $\chi=0$, for a long time during which the spacetime expands exponentially with Hubble rate
  \be H_I = \frac{\sqrt{\bar\beta} \chi_0^2}{4\sqrt{3}\bp},\label{HI}\ee
  where $\bp$ is the reduced Planck mass. Eq.~(\ref{HI}) can be obtained by using the trace of the Einstein equations and  noting that the potential energy density difference between the false and the true vacuum is, according to Eq.~(\ref{CWpot}), $\bar\beta\chi_0^4/16$. After this exponential expansion, $\chi$ goes towards the true minimum, $\chi_0$, and reheating occurs up to a temperature $T_r$ at most given by 
  \be T_r^4 \approx \frac{15 \bar\beta \chi_0^4}{8\pi^2 g_*(T_r)},\label{TRHmax}\ee
  where $g_*(T)$ is the effective number of relativistic species at temperature $T$.
  
  The amount of supercooling that actually takes place depends on the model and can be parameterized by~\cite{Salvio:2023qgb}
  \be \epsilon\equiv  \frac{g^4}{6\bar\beta \log\frac{\chi_0}{T}},
 \label{CondConv}\ee
where $g$ plays the role of a ``collective coupling" of $\chi$ with all fields of the theory: in the model of Sec.~\ref{A minimal model}, 
\be g=\sqrt{3}|e_d|, \qquad \bar\beta = \frac{6 e_d^4}{(4\pi)^2}=\frac{2g^4}{3(4\pi)^2}. \label{gbeta}\ee 
The larger the supercooling, $T\ll \chi_0$, the smaller $\epsilon$. 

As shown in~\cite{Salvio:2023qgb} and~\cite{Salvio:2023ynn}, if $\epsilon\ll 1$,  one can describe the PT, as well as the GW spectrum and the PBH production, with good accuracy by means of a small-$\epsilon$ expansion (called ``supercool expansion"). If $T_r$ is given by~(\ref{TRHmax}), at leading order (LO) one can parameterize the relevant physics in terms of $\chi_0$, $g$, $\bar\beta$  and $g_*$ only, while at next-to-leading order (NLO) one must introduce an extra parameter $\tilde g$, which in the model of Sec.~\ref{A minimal model} is $\tilde g = g/\sqrt[6]{3}$.
  At LO the nucleation temperature $T_n$ (defined as the temperature at which the false-vacuum decay rate per unit volume $\Gamma_v$ equals $H_I^4$),  can be computed with the simple formula
\be T_n\approx \chi_0\exp\left(\frac{\sqrt{c^2-16a}-c}8\right), \label{appTn}\ee
with
\be a\equiv \frac{c_3g}{\sqrt{12}\bar\beta},\quad c_3=18.8973... \quad c\equiv 4\log\frac{4\sqrt{3}\bp}{\sqrt{\bar\beta}\,\chi_0}+\frac32\log\frac{a}{2\pi}. \nonumber\label{caDef}\ee
 
 If one wants to describe not only models with $\epsilon\ll1$, but also those with $\epsilon$ of order 1, one can again describe the relevant physics perturbatively in the amount of supercooling, but through an improved version of the supercool expansion~\cite{Salvio:2023ynn}, where one has to exactly take $\tilde g$ into account at any order~\cite{Salvio:2023ynn}.   In this ``improved supercool expansion" $T_n$ can be computed with a more complex procedure explained in~\cite{Salvio:2023ynn}.

\begin{figure}[t!]
\begin{center} 
\hspace{-0.37cm} 
\vspace{0.4cm}
 \includegraphics[scale=0.44]{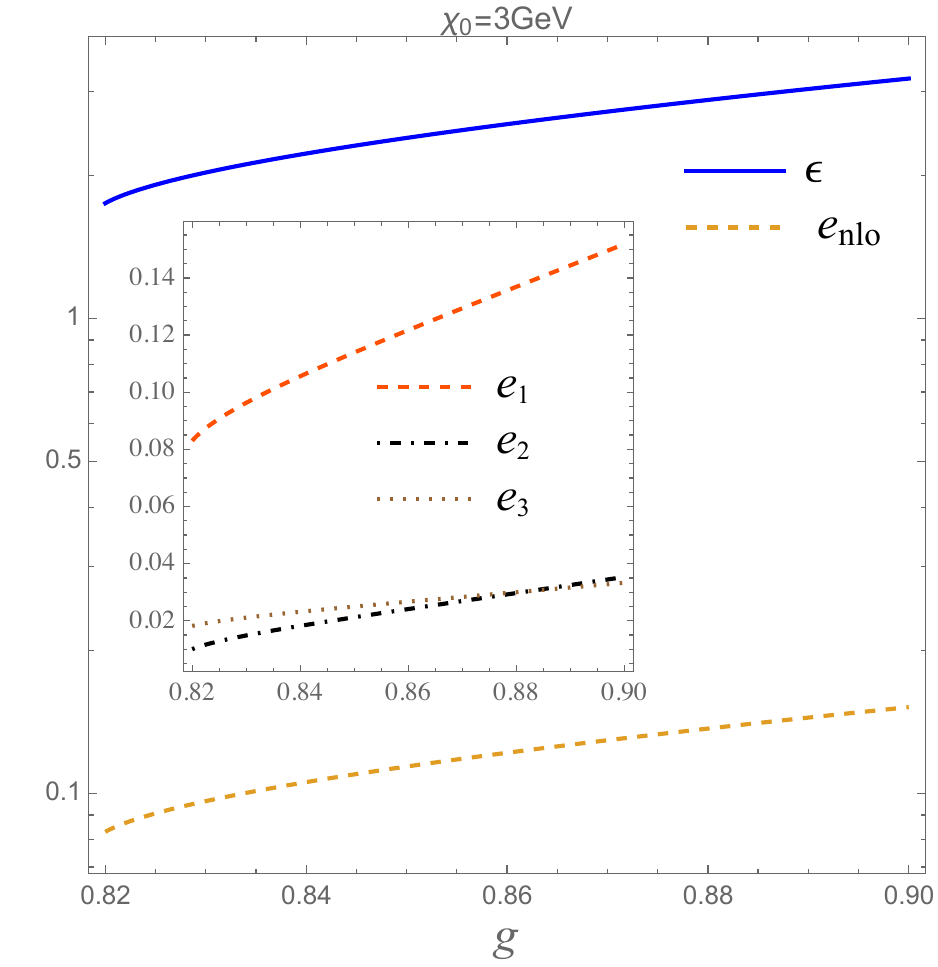} 
 \end{center} 
 \vspace{-0.5cm}
   \caption{\em Comparison between $\epsilon$ and the error $e_{nlo}$  that one is making in using the standard supercool expansion at NLO in the minimal model of Sec.~\ref{A minimal model}. The inset shows the three sources of errors defined in~(\ref{enlo}). Varying $\chi_0$ around the GeV scale does not show significant differences: a plot corresponding to another value of $\chi_0$ is shown in the appendix.}
\label{SE}
\end{figure}

Let us now quantify the error that one is making in analysing the model of Sec.~\ref{A minimal model} with the standard supercool expansion at NLO. Fig.~\ref{SE} shows $\epsilon$ as well as $e_{nlo}\equiv \max(e_1,e_2,e_3)$ and $(e_1,e_2,e_3)$ defined by
\be (e_1,e_2,e_3)\equiv \left(\frac{\tilde c_3^2 \tilde g^6 \epsilon}{2\pi^2 c_3^2 g^6}, \frac{|\frac12\log(\epsilon/g^2)-1/4|}{\log\frac{\chi_0}{T_n}},\frac{\epsilon}{96}\right) \label{enlo}\ee
(computed for simplicity with the LO formula for $T_n$ in~(\ref{appTn})) with $\tilde c_3= 31.6915 ...$\,. As discussed in~\cite{Salvio:2023ynn}, $e_1$ quantifies the above-mentioned error, while $e_2$ and $e_3$ are estimates of the errors present also in the improved supercool expansion at leading order.
 Fig.~\ref{SE} tells us that such improved supercool expansion is valid and a better approximation than the standard supercool expansion  as $\epsilon\sim1$ and $\{e_1>e_2, e_1> e_3\}$ in this model. Therefore, in the rest of the paper we will use the improved supercool expansion.

In general the existence of $T_n$ restricts the parameter space~\cite{Salvio:2023ynn}. One might wonder if the PT always complete, especially in the region of the parameter space where $T_n$ does not exist. If $\Gamma_v$ is not large enough to ensure the completion of the PT the temperature keeps decreasing and eventually the effect of the spacetime curvature, as well as quantum fluctuations, can become important in the decay rate~\cite{Kearney:2015vba,Joti:2017fwe,Markkanen:2018pdo,DelleRose:2019pgi}. This can lead to the completion of the transition. Indeed, during the exponential expansion, the non-minimal coupling in~(\ref{non-min}), with   $\xi_\chi$ positive, gives a tachyonic mass contribution, $ -6\xi_\chi H_I^2\chi^2$, to the effective potential of $\chi$, which makes the transition complete  when, eventually, the spacetime curvature dominates over thermal fluctuations. This contribution allows the field to reach the absolute minimum of the effective potential and ensures that percolation (i.e.~when most of the vacuum energy is converted into radiation, see Sec.~\ref{Preheating and reheating}) occurs. We, therefore, assume that $\xi_\chi$ is positive at least when this mechanism is needed to complete the PT.
  
  \section{Preheating and reheating}\label{Preheating and reheating}
  \vspace{-0.2cm}
 
 Can one has a sufficiently high reheating temperature $T_r$ in this model through direct decays of   $\chi$ into SM particles? The answer to this question seems negative for the following reason. 

The only tree-level channels for the decay of $\delta\chi$ into SM particles are $\chi\to A'A'(HH)\to$ 4 fermions, where the intermediate $A'$s and Higgs bosons $H$ are virtual because an on-shell $A'$ or $H$ is much heavier than $\delta\chi$ in our setup. The decays of $A'$ and $H$ into fermions are allowed by the interaction with the electromagnetic current in~(\ref{mdIntApSM}) and the Yukawa couplings, respectively. Dropping the external legs, the amplitudes for these processes for light-enough leptons are of order $e_d^2\chi_0 (e\eta)^2/m_d^4(\chi_0)$ and $\lambda_{\chi h}\chi_0 y_{\rm SM}^2/M_h^4$, respectively, where the denominators $m_d^4(\chi_0)$ and $M_h^4$ are due to the low energy limit of the two $A'$ or  $H$ propagators
and $y_{\rm SM}$ is the SM Yukawa coupling of a light SM fermion. Taking the absolute value squared of the amplitude and integrating over the four-body phase space to obtain the decay rate $\Gamma$ one finds 
\be \Gamma(\chi\to A'A'\to 4~\mbox{leptons}) \lesssim m_\chi^7\left(\frac{e_d^2\chi_0 (e\eta)^2}{m_d^4(\chi_0)}\right)^2, \label{Ch1}\ee
\be \Gamma(\chi\to HH\to 4~\mbox{leptons}) \lesssim m_\chi^7\left(\frac{\lambda_{\chi h}\chi_0 y_{\rm SM}^2}{M_h^4}\right)^2 \label{Ch2}\ee
or even smaller because of four-body phase-space suppressions. The factor $m_\chi^7$ has been inserted for dimensional reasons.
As usual the scattering channels (such as $\chi\chi\to A'A'(HH)$) are not efficient in converting the energy density of $\chi$ into SM radiation density because  the inverse scattering processes ($A'A'(HH)\to\chi\chi$) balance them and, therefore,  the net average energy transfer vanishes. Moreover,
setting $\eta$ compatible with the experimental limits, $\chi_0\lesssim 10$~GeV and $g$ to realistic values compatible with the PTA GW signal, one finds that 
the channels in~(\ref{Ch1})-(\ref{Ch2}) can reheat the universe only up to a reheating temperature below  the MeV scale, too small to realize a sufficiently fast reheating, where $T_r$ is rather given by the last equation in~(\ref{TRHmax}).

Let us then consider the preheating phase~\cite{Dolgov:1989us,Traschen:1990sw,Kofman:1994rk,Kofman:1997yn}, when production of particles interacting with $\chi$ occurs as a result of the time dependence of this field through parametric resonance. In this model the only particle  with a sizable coupling with $\chi$ is $A'$ and its field equations are
\be \frac1{\sqrt{|\det \hat g|}}\partial_\nu\left(\sqrt{|\det \hat g|}\, F'^{\mu\nu}\right)= m_d^2(\chi) A'^\mu+e_d J'^\mu -e\eta J^\mu, \label{ApEq}\ee
where $\det \hat g$ is the determinant of the spacetime metric, $F'_{\mu\nu}\equiv \partial_\mu A'_\nu-\partial_\nu A'_\mu$ and we have used the smallness of $\eta$. After the false vacuum decay has taken place mostly of the energy stored in $\chi$ is thus transferred to $A'$ because  $H_I^4$ is tiny compared to the potential density in the setup under study.

At some stage $\chi$ undergoes small oscillations around $\chi_0$. In such a  regime one can  analytically show that most of the energy density stored in $\chi$ is efficiently transferred  to $A'$ as we now do.
 In that case, the dependence of $\chi$ on the cosmic time $t$ is given by 
\be \chi(t) =\chi_0+\Phi \sin(m_\chi t), \label{SmallOsc}\ee
 with $\Phi\ll\chi_0$, and the background dependent mass $ m_d(\chi)\sim e_d\chi_0$ is generically 
extremely large compared to $H_I$ (see Eq.~(\ref{HI})).  
So we can 
neglect the expansion of the universe.
Recalling that $\eta$ must be very small and setting the unitary gauge, where $J'_\mu=0$, 
Eq.~(\ref{ApEq}) then reads
\be  \partial_\nu\, F'^{\mu\nu}= m_d^2(\chi) A'^\mu. \label{ApEOMosc}\ee
Taking the divergence of this equation 
and inserting the result back into Eq.~(\ref{ApEOMosc}) gives
\be -\partial_\nu\partial^\nu A'_\mu=m_d^2(\chi) A'_\mu +\partial_\mu\left(\frac{A'^\nu \partial_\nu m_d^2(\chi)}{m_d^2(\chi)}\right).\ee
In the regime (\ref{SmallOsc}) with $\Phi\ll\chi_0$ the second term in the right-hand side of the equation above is negligible with respect to the first one because suppressed by extra powers of $\chi_0$ in the denominator and of $m_\chi = \sqrt{\bar\beta}\chi_0\ll \chi_0$ in the numerator:
\be \bac  \left\{\bac m_d^2(\chi) A'_\mu\approx e_d^2(\chi_0^2+2\chi_0\Phi \sin(m_\chi t))A'_\mu \\
\partial_\mu\left(\frac{A'^\nu \partial_\nu m_d^2(\chi)}{m_d^2(\chi)}\right)  \approx  \partial_\mu\left(\frac{A'_0 2\chi_0\Phi m_\chi \cos(m_\chi t)}{\chi_0^2}\right)\ea \right. .
 \ea \ee
So the field equations of $A'$ simply reduce to that of a scalar  with a quartic portal interaction with $\chi^2$. Working in the Fourier space of momenta $\vec k$, all three polarizations $A'_k$ of the dark photons are then well described by the equation
\be \ddot A'_k + (k^2+e_d^2\chi_0^2+2e_d^2\chi_0\Phi\sin(m_\chi t))A'_k = 0, \ee
with $k\equiv|\vec k|$. This equation  can now be brought into the well-known Mathieu equation:
\be \frac{d^2 A_k'}{dz^2} + (a_k-2q \cos(2z))A'_k = 0, \label{MatAkp}\ee 
where $a_k\equiv 4(k^2+e_d^2\chi_0^2)/m_\chi^2$, $q\equiv 4e_d^2\chi_0\Phi/m_\chi^2$ and $z\equiv m_\chi t/2+\pi/4$. Therefore, as explained in~\cite{Lachlan,Kofman:1997yn} (see also Ref.~\cite{Armendariz-Picon:2007gbe} for a more recent analysis), there is an exponential growth of the dark photon field, $A_k' \sim \exp(\mu_k^{(n)}z)$, which corresponds to an equally explosive growth of dark photon occupation numbers of quantum fluctuations when $k$ belongs to  some resonance bands labelled by an arbitrary integer $n$. These ranges of values of $k$ always exist and the constants of the exponential growth, the $\mu_k^{(n)}$, allow to efficiently transfer  all the energy density stored in the inflaton to the dark photon in a negligible cosmic time. This exponential growth cannot last forever because of the backreaction of $A'$ on $\chi$: when $A'$ grows $\chi$ feels a growing effective mass, which makes $\chi$ tend to a vanishing value as time passes.

However, the preheating phase allows to transfer the vacuum energy density $\rho_V$ to $A'$, which can then decay into pairs of SM fermions, thanks to the interaction with 
$J_\mu$ in~(\ref{mdIntApSM}). The corresponding tree-level decay rate into lepton pairs is 
\be \Gamma(A'\to 2~\mbox{leptons}) = \frac{(\eta e)^2 m_d(\chi_0)}{12\pi} \Phi_s(m_d(\chi_0),m_l), \ee
where $m_l$ is the  lepton mass and the phase space factor $\Phi_s$ is
\be \Phi_s(m_d,m_l) = \sqrt{1-\left(\frac{2m_l}{m_d}\right)^2}\left(1+\frac{2m_l^2}{m_d^2}\right) \theta(m_d-2m_l).\ee 
Unlike for the decays of $\delta\chi$ into four SM fermions, which we have previously discussed, setting $\eta$ compatible with the experimental limits, $\chi_0\lesssim 10$~GeV and $g$ compatible with the PTA GW signal and all observational constraints, one finds that these processes allow for a very fast reheating.
Including the hadronic decay channel of $A'$ reinforces this result. Because $\rho_V$ represents the full energy budget of the system, the value of $T_r$ given by Eq.~(\ref{TRHmax}) is the correct value of the reheating temperature in this model. Therefore, given $g$, $\chi_0$ and $g_*(T_r)$, the temperature $T_r$ can be computed through Eq.~(\ref{TRHmax}):
\be T_r = \frac{ \sqrt[4]{\frac{5}{g_*(T_r)}}\, g\, \chi_0}{2 \sqrt{2} \pi },\label{Trsimp}\ee
where the expression of $\bar\beta$ in terms of $g$ in Eq.~(\ref{gbeta}) has been used.

  \vspace{-0.1cm}
  
 \section{Gravitational waves and primordial black holes}\label{Gravitational waves and primordial black holes}
 
 In the improved supercool expansion one can also obtain accurate expressions for $\beta/H_n$, where $\beta$ is the inverse duration of the PT and $H_n$ is the Hubble rate at $T=T_n$, as well as the GW spectrum and its red-shifted frequency peak today $f_{\rm peak}$~\cite{Salvio:2023ynn}. Moreover, this expansion can be used to describe the PBH late-blooming  mechanism in terms of the few parameters $\chi_0$, $g$, $\bar\beta$ and $\tilde g$~\cite{Salvio:2023ynn}. We now make use of these results  in the context of the minimal model of Sec.~\ref{A minimal model}.
 
  \begin{figure}[t!]
\begin{center} 
  \includegraphics[scale=0.49]{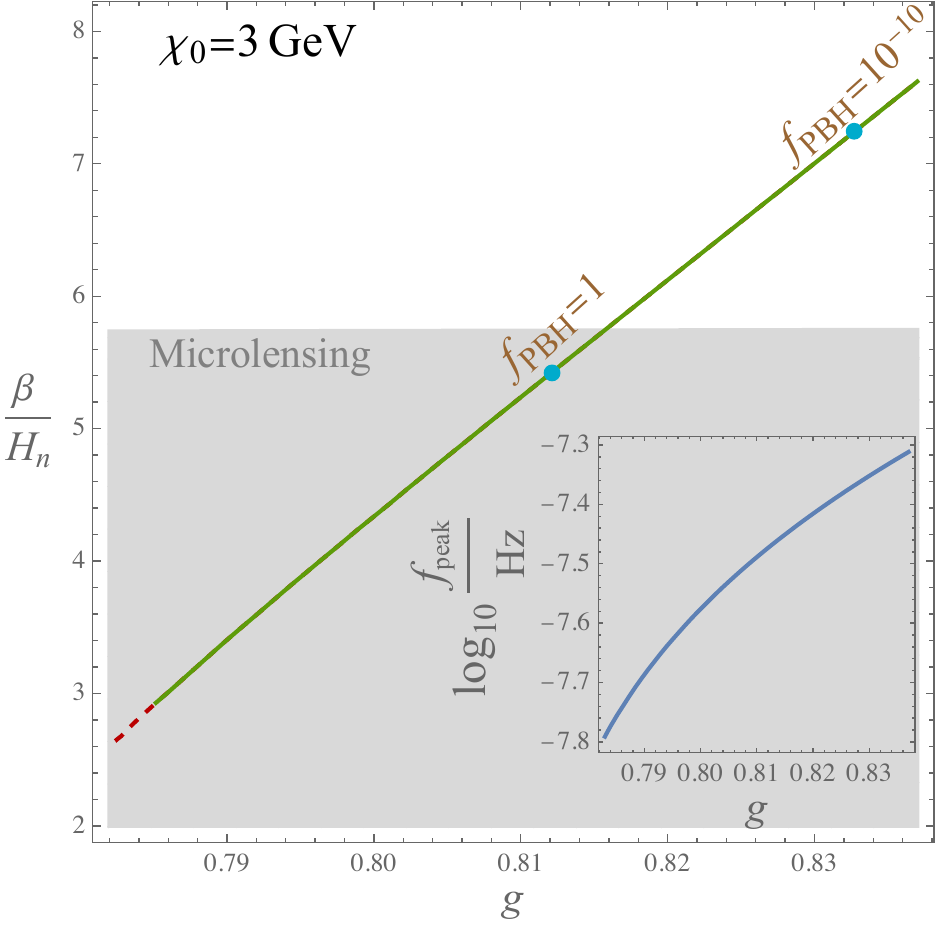} \\
  \vspace{0.4cm}
  \includegraphics[scale=0.53]{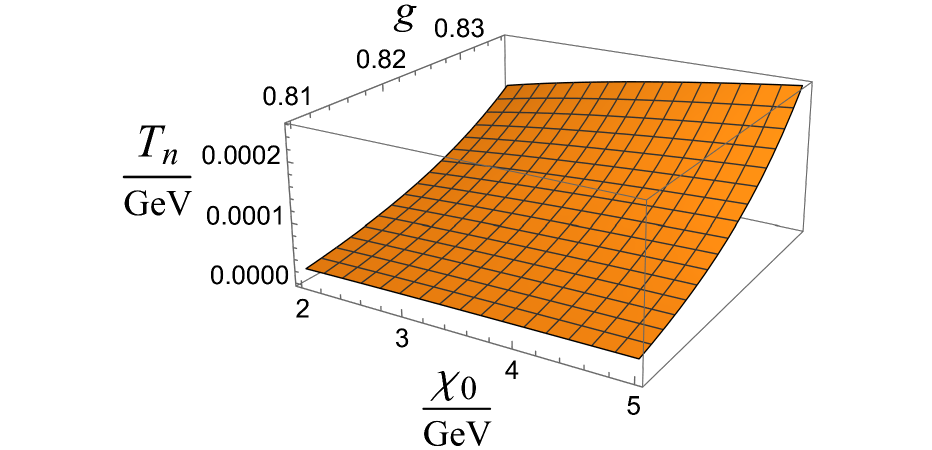}  
    \caption{\em {\bf Upper plot.}  $\beta/H_n$, and (in the inset) $f_{\rm peak}$ in the minimal model of Sec.~\ref{A minimal model} and computed with the improved supercool expansion. Here $g_*(T_r)=6$, although the plot depends weakly on $g_*$. Values of $g$ corresponding to the green solid line reproduce the NANOGrav signal at 68\% C.L.; all displayed  values of $g$ correspond to $\epsilon$ at most of order 1, so the improved supercool expansion can be trusted. The shaded region is constrained by microlensing. {\bf Lower plot.} $T_n$ as a function of $\chi_0$ and $g$ in the improved supercool expansion.}\label{betafPDS}
  \end{center}
\end{figure}

 Fig.~\ref{betafPDS}  shows the predictions of this model for $\beta/H_n$, $f_{\rm peak}$ and $T_n$ setting $\chi_0$ at the GeV scale.  The plot shows that there are values of $g$ such that this model reproduces the GW background detected by PTAs and is not excluded by other observations. As explained in~\cite{Gouttenoire:2023naa}, $f_{\rm PBH}$ can be computed once the temperature in~(\ref{Trsimp}) and $\beta/H_n$ are known. In our model, these two quantities are functions of $g$ (for given $\chi_0$ and $g_*$). In particular $\beta/H_n$ decreases very rapidly when $g$ approaches from above a critical value which is around 0.8. This is the manifestation of a general model-independent feature of phase transitions in the presence of RSB: when one approaches a critical line in the $\{g,\bar \beta\}$ plane (at fixed $\tilde g$ and $\chi_0$) $\beta/H_n$ decreases very rapidly (see Figs.~3 and 4 of Ref.~\cite{Salvio:2023ynn}). On the other hand, in~\cite{Gouttenoire:2023naa} it was shown that when $\beta/H_n$ decreases the fraction $f_{\rm PBH}$ increases exponentially in this scenario. Therefore, $f_{\rm PBH}$ is a function of $g$ (for fixed $\chi_0$ and $g_*$) that is very rapidly growing when $g$ approaches the critical value from above. In particular the values of $g$ corresponding to $f_{\rm PBH}=1$ and $f_{\rm PBH}=10^{-10}$ and the microlensing constraints~\cite{MACHO:2000qbb,EROS-2:2006ryy,Niikura:2019kqi,Niikura:2017zjd} (the strongest constraints on PBHs in the setup of the upper plot in Fig.~\ref{betafPDS}) are given in Fig.~\ref{betafPDS}.  These PBHs weight about a solar mass~\cite{Gouttenoire:2023bqy}.

In the appendix we provide the corresponding plots for other values of $\chi_0$ around the GeV scale. These have the same qualitative feature. Also, as expected,  they show  that when $\chi_0$ goes above the GeV scale the values of  $g$ that can reproduce the PTA GW background tend to disappear. Moreover, note that using the values of $g_*(T_r)$ in the SM (see e.g.~\cite{Borsanyi:2016ksw}), Eq.~(\ref{Trsimp}) leads to a reheating temperature of order $10^2$~MeV  for the realistic values of the parameters, which explain the background of GWs found by PTAs.

Furthermore, although PBHs are produced, one cannot account for the  entire dark matter abundance through PBHs, as clear in the upper plot of Fig.~\ref{betafPDS} and in Fig.~\ref{betafPDSA}. Also, the dark photon in this model cannot contribute significantly to dark matter: as evident in Fig.~\ref{betafPDS} (upper plot) and Fig.~\ref{betafPDSA}, $g$ should be of order one, which leads, using $\chi_0\sim$~GeV, to an almost GeV-scale dark photon, not to a sub-MeV one~\cite{Fabbrichesi:2020wbt}. One can complete the dark matter abundance as well as reproduce the observed neutrino oscillations and matter antimatter asymmetry by introducing, for example, three right-handed neutrinos along the lines of Refs.~\cite{Dodelson:1993je,Akhmedov:1998qx,Shi:1998km,Abazajian:2001nj,Asaka:2005pn,Canetti:2012vf,Koutroulis:2023fgp}. Since neutrinos are not charged under $U(1)_D$ and do not appear in the electromagnetic current, the preheating and reheating analysis of Sec.~\ref{Preheating and reheating} remains valid.

  \vspace{-0.1cm}
   
\section{Conclusions}\label{Conclusions}
  \vspace{-0.2cm}
We have constructed and studied an explicit realistic model featuring RSB (and thus an associated supercooled PT) that explains the background of GWs found by PTAs. The model features a simple dark sector, which involves the field $\chi$ responsible for RSB and a hidden photon that can interact with the SM particles through a mixing with the SM photon. The completion of the transition is ensured by a non-minimal coupling between $\chi$ and $R$. Moreover, fast reheating after the supercooled PT occurs thanks to a preheating phase, where the vacuum energy density stored in $\chi$ is transferred to the dark photon, which  can then decay to SM fermions via the mixing. Finally, we have also identified regions of the parameter space where one has significant PBH production.

The work presented here opens several research directions. As an outlook example, it would be interesting to generalize the dark sector: the model presented here is the minimal RSB one able to reproduce the GW background detected by PTAs, but one can introduce other scalar, vector and fermion fields with sizable couplings to $\chi$.  Moreover, it would be valuable to extend the preheating analysis reported here to a general PT associated with RSB in order to identify in a model-independent way the parameter space corresponding to fast reheating.
Finally, the interaction between the theoretical and observational community (PTAs, the LIGO and VIRGO collaborations for GW detection, etc.) will be vital to further probe this scenario.

\vspace{-0.2cm}

\section*{Acknowledgments} 
\vspace{-0.1cm} 

\noindent    This work has been partially supported by the grant DyConn from the University of Rome Tor Vergata.

\vspace{0.cm}

\appendix

\section{\\ Varying the symmetry breaking scale}\label{appendix}

In this appendix the plots in Figs.~\ref{SE} and~\ref{betafPDS} (upper plot) are extended to other values of the symmetry breaking scale $\chi_0$ . This is done to illustrate the quantitative dependence on  $\chi_0$.

\begin{figure}[t!]
\begin{center} 
\hspace{-0.37cm} 
\vspace{0.4cm}
 \includegraphics[scale=0.41]{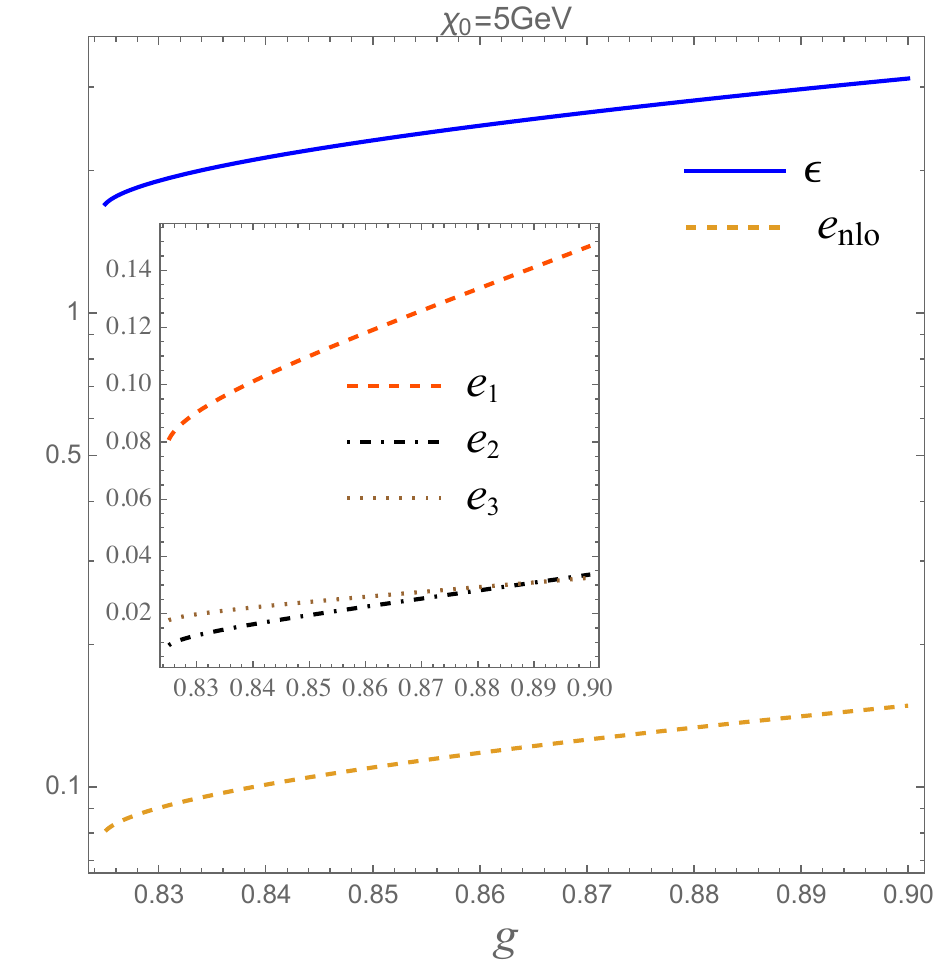} 
 \end{center} 
 \vspace{-0.5cm}
   \caption{\em Like in Fig.~\ref{SE}, but with $\chi_0=5GeV$.}
\label{SEA}
\end{figure}

Fig.~\ref{SEA} shows that varying $\chi_0$ around the GeV scale does not significantly change the plot in Fig.~\ref{SE}.

\begin{figure}[t!]
\begin{center}  
\includegraphics[scale=0.39]{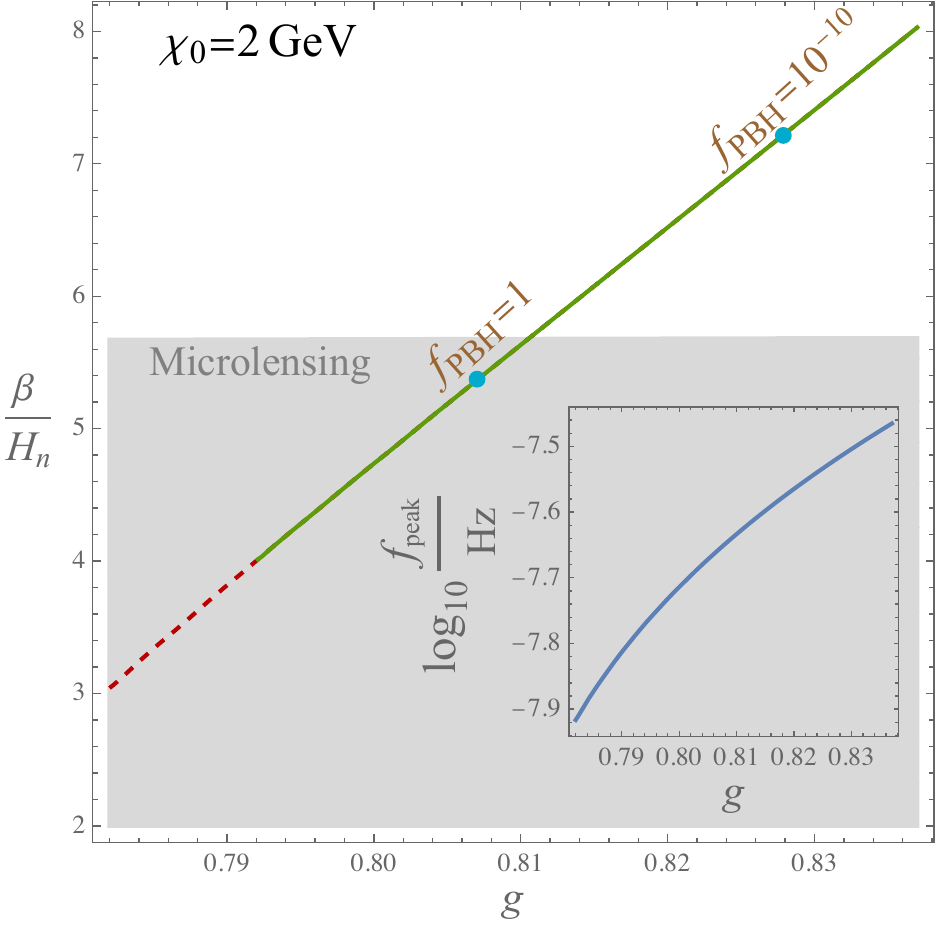}  
\hspace{1cm} \\
   \includegraphics[scale=0.39]{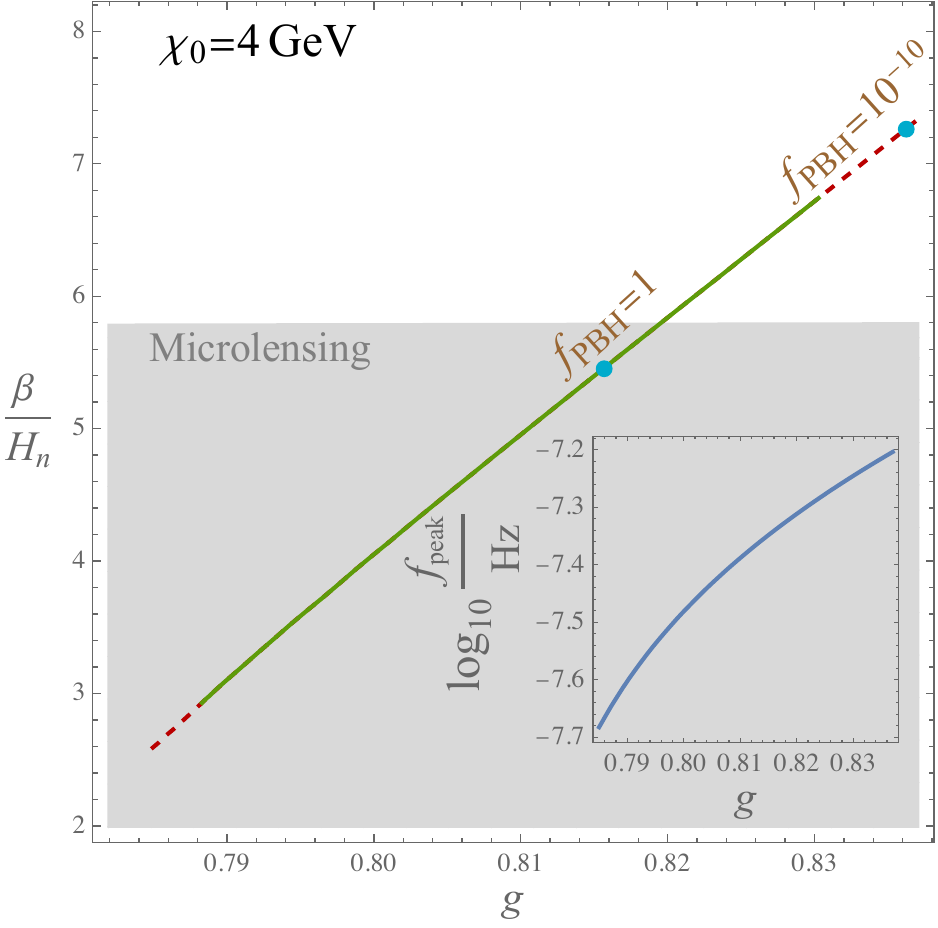}  \hspace{1cm} \\ \includegraphics[scale=0.39]{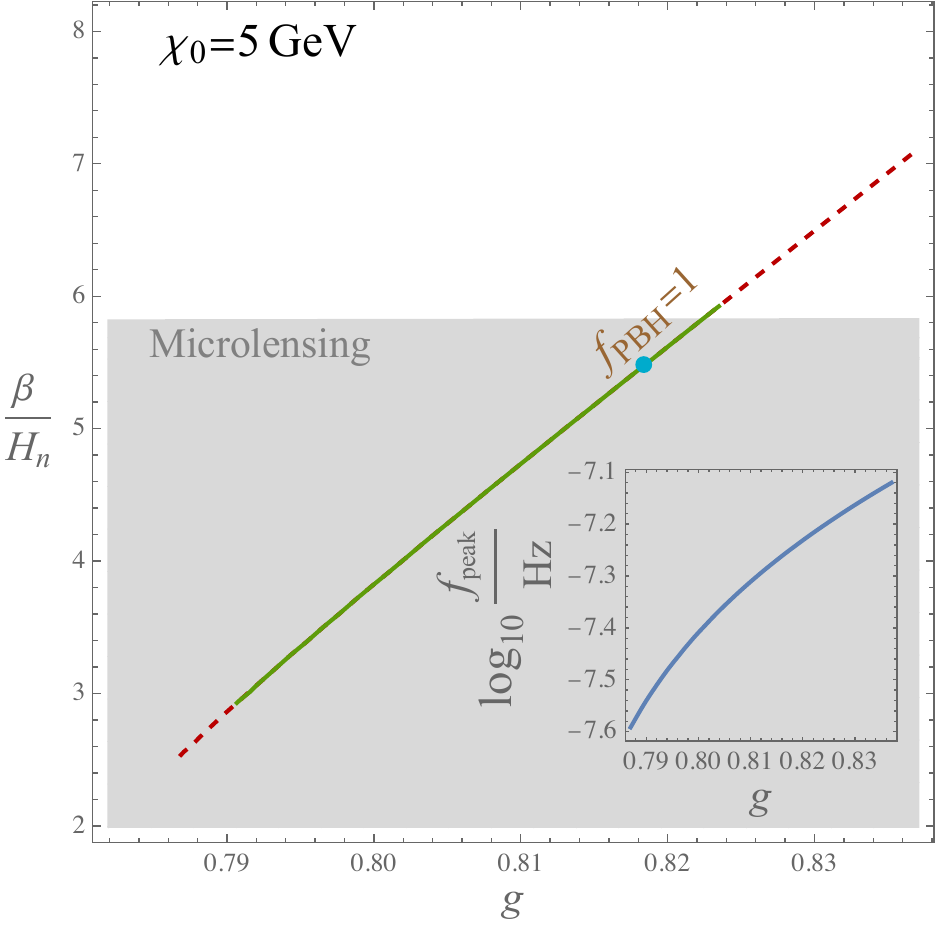}
    \caption{\em Like in the upper plot of Fig.~\ref{betafPDS}, but for other values of $\chi_0$ around the GeV scale.}\label{betafPDSA}
  \end{center}
\end{figure}

Fig.~\ref{betafPDSA} shows that varying $\chi_0$ around the GeV scale in Fig.~\ref{betafPDS} (upper plot) leads to qualitatively similar plots, apart from the expected fact that when $\chi_0$ goes above the GeV scale, the values of  $g$ that can reproduce the PTA GW background tend to disappear.

\end{document}